\documentclass[ floatfix,secnumarabic, amssymb, nobibnotes, aps, prd]{revtex4-2}

\usepackage{mathrsfs}

\usepackage{amssymb,mathtools,amsmath}

\usepackage{graphicx, bm}

\begin{document}

\title{ The effect of Quantum Time Crystal Computing to Quantum Machine Learning methods
 }%
 
\author{Hikaru Wakaura}%
\email[Quantscape: ]{
hikaruwakaura@gmail.com} 
\affiliation{QuantScape Inc. QuantScape Inc., 4-11-18, Manshon-Shimizudai, Meguro, Tokyo, 153-0064, Japan}
  
\author{Andriyan B. Suksmono} 
 
\affiliation{ The School of Electrical Engineering and Informatics, Institut Teknologi Bandung (STEI-ITB), Jl. Ganesha No.10, Bandung, Indonesia }%
   
 \affiliation{ Research Collaboration Center for Quantum Technology 2.0, BRIN-ITB-TelU, Indonesia }

 \affiliation{ ITB Research Center on ICT (PPTIK-ITB) }

\email[Bandung Institute of Technology: ]{suksmono@itb.ac.id}

\begin{abstract}    

Many body localization shows the robustness for external perturbations and time reversal symmetry on Time Crystal.
This Time Crystal  prolongs the coherence time, hence, it is used for quantum computers as qubits. 
Therefore, we established the method to exploit Time Crystals for quantum computing by controlling external noise called Quantum Time Crystal  Computing and demonstrated solving the problem of generating correct waves using Quantum Reservoir Computing, and fitting of given function using Quantum Neural Network and Variational Quantum Kolmogorov-Arnold Network.
 As a consequence, we revealed that Quantum Time Crystal  Computing lower the accuracy of Quantum Reservoir Computing and improved the accuracy of Quantum Neural Network and Variational Quantum Kolmogorov-Arnold Network. 
This result may be the one of milestones of Quantum Error Mitigation as the case that noise improves the accuracy of Quantum Machine Learning.

\end{abstract}        
\maketitle  
 \tableofcontents

\section{Introduction}\label{1}

 Quantum computation, with its potential to solve complex problems exponentially faster than classical computers, represents a revolutionary paradigm in computational science. However, the practical realization of large-scale quantum computers faces a significant challenge: quantum systems are inherently fragile and prone to errors due to decoherence, noise, and imperfections in quantum gates. Overcoming these errors is essential to harness the full potential of quantum computing. Two key approaches have emerged in response to this challenge: Quantum Error Correction (QEC) and Quantum Error Mitigation (QEM).   
Quantum Error Correction, first developed in the 1990s, is a rigorous framework designed to protect quantum information from errors by encoding logical qubits into entangled states of multiple physical qubits. By detecting and correcting errors at the physical level, QEC enables the possibility of fault-tolerant quantum computation, where the logical computation remains stable despite the presence of physical errors. The most widely studied quantum error-correcting codes, such as the surface code \cite{PhysRevA.85.060301} \cite{acharya_suppressing_2023} and the Shor code, \cite{georgescu_25_2020} have provided a foundation for scalable quantum computing architectures. However, QEC comes at the cost of substantial overhead in terms of the number of qubits and the complexity of operations, which has slowed its immediate implementation in near-term quantum devices. 
In contrast, Quantum Error Mitigation represents an alternative approach better suited for today’s noisy intermediate-scale quantum (NISQ) devices. Rather than preventing errors at the hardware level, QEM focuses on reducing the impact of errors at the software or algorithmic level, enabling accurate computation without the need for full-scale error correction. Techniques such as extrapolation \cite{PhysRevX.8.031027}, probabilistic error cancellation, and noise-aware post-processing such as Quantum Subspace Expansion \cite{2022PhRvL.129b0502Y} \cite{Cai2021quantumerror} are designed to minimize the effect of errors on the final output of a quantum algorithm. While QEM is not a replacement for QEC in the long term, it offers a pragmatic solution for improving the accuracy of quantum computations on current devices, where error rates remain high, and fully fault-tolerant quantum computation is not yet feasible.  
Before the century, integrable \cite{2007arXiv0706.1579L} and many body localized quantum systems \cite{PhysRevLett.114.140401} \cite{PhysRevB.88.014206} has been fostered. 
These systems exhibit the weak eigenstate thermalization hypothesis, robustness for external perturbations, and unique behavior to noise. 
They fascinate scientists in not only quantum statical mechanics but also quantum informational physics, and emulating and exploiting are endeavored by many groups.
For these reason,chaotic phenomena are exploited for conputing \cite{TEH2020102421} such as image encription \cite{math11112585}. 
Quantum chaos is also used for computing by various ways. 
Quantum Reservoir Computer is the major example that uses chaotic behavour \cite{2024arXiv240612948C} \cite{2024OExpr..3228441Z} emerged by measurement and feedback \cite{2024arXiv240615783K} . 
The circuit to emerge quantum chaos \cite{2024arXiv240513892S},two-level systems of chaos \cite{2024PhDT.........4K},and Shor algorithm using quantum integrable systems \cite{2024PhRvR...6c2046P} are also studied.

Time Crystals (TCs), first introduced by Nobel laureate Frank Wilczek in 2012 \cite{2012PhRvL.109p0401W}, mark a significant departure from conventional understandings of equilibrium dynamics. Unlike traditional systems where time symmetry is maintained, TCs exhibit a spontaneous breaking of time-translation symmetry, leading to persistent oscillations even without external driving forces. This remarkable property opens up new avenues for exploring novel phases of matter, with exciting potential applications in quantum computing and precision metrology.
Several research groups around the world have demonstrated that Discrete Time Crystals \cite{2020ARCMP..11..467E, 2020arXiv201205495C} can extend the lifetime of qubits. Additionally, researchers at Aalto University have shown that TCs can be used as qubits, maintaining coherence for over 2 seconds after energy level crossings \cite{autti_nonlinear_2022}.
Although gate operations break time-reversal symmetry, controlling the quantum state through the TC phase may still be possible. Noise modulation can propagate the quantum state toward a target state while preserving time-reversal symmetry. As a result, we propose a new approach to quantum computation, which we call Quantum Time Crystal Computing (QTCC)\cite{QTQT}, that leverages the unique properties of Time Crystals.
 
The optimization of parameter and demonstration of quantum calculation are revealed to be possible, thus, this paper show the result of demonstration of Quantum Machine Learning (QML) \cite{2019QS&T....4a4001K,2019Natur.567..209H}. 
QML is the hybrid method using a quantum computer and classical machine learning methods, such as Quantum Circuit Learning (QCL) and Quantum Support Vector Machine (QSVM) \cite{2014PhRvL.113m0503R}, Quantum Neural Network (QNN) \cite{2022PhRvA.106b2601A,2022arXiv220211200K}, Born Machine (BM) \cite{2018arXiv180404168L},and Variational Quantum Eigensolver's family \cite{Kassal2011,McClean_2016,Grimsley2019,2019arXiv190608728P,2021arXiv210501141W,2021arXiv210902009W}. 
In detail this paper show our result of Quantum Reservoir Computing (QRC) Quantum Neural Network (QNN) and Variational Quantum Kolmogorov-Arnold Network (VQKAN) \cite{Wakaura_VQKAN_2024,2025arXiv250321336W,2025arXiv250322604W} by QTCC. 
   
 As a result, we revealed that QTCC is able to process tasks more accurately than the systems manipulated by noise-less external coherent electromagnetic field.       
 This result will be the one of the milestones of Quantum Error Mitigation as the case that the noise improves the accuracies of QML.   
         
Section \ref{1} is the introduction, section \ref{2} describes the method detail of QTCC and the optimization method, section \ref{3} describes the result of demonstration of echo using QRC by QTCC, and fitting using QNN and VQKAN by QTCC, \ref{4} is the concluding remark.   
   
\section{Method}\label{2} 
 This section provides a detailed explanation of Quantum Time Crystal Computing (QTCC) and our experimental demonstration. A Time Crystal (TC) is a system where quantum states propagate symmetrically with respect to time-reversal symmetry. Such behavior is observed in several quantum Floquet 
 systems, including a ring of spins and photons. 
 In these systems, the Hamiltonian alternates periodically between one driven by an electromagnetic field and another representing noise, modeled by a Hamiltonian with randomized coefficients corresponding to the noise. Any quantum system can enter the TC phase, and being in this phase extends the system's lifetime, as shown by the equation:

\begin{equation} 
 H = 
 \begin{cases}
 \sum_{j = 0}^{N} 0.5 (1-d) X_j & \text{ $ (0 \leq t~ mod ~ 2 T \leq T) $ } \\ 
 H_1 & \text{ $ (T < t ~ mod ~ 2 T\leq 2T). $ } \\ 
 \end{cases}
\end{equation}

$N$ is the number of qubits and $ H_1 $ includes the noises and $ d $ is the assumed tunable coefficient, which is 0.001 for all calculations. 
 We assume the Hamiltonian All-to-all connected Ising model Hamiltonian the noise of this term as the rotation noise of parameters such as $ \{\theta_i = \theta_i ^0 + \theta_i ^r Err(0, 1 /3) \} $, Where $ Err (0, 1 /3) $ is the Gaussian function which the center is $ x = 0 $and standard deviation is 1 /3. 

We demonstrated the Quantum Resavoir Computing (QRC), and fitting problem using Quantum Neural Network (QNN), Variational Quantum Kolmogorov-Arnold Network (VQKAN) \cite{Wakaura_VQKAN_2024} by QTCC. 
QRC is one of the QML methods that makes the filter matrix from input data and encoded result of time propagation of quantum system.
 This method can learn and be able to predict from data without optimization method. 
 We assume the filter matrix $ W $ as, 
 
\begin{equation} 
W = V ^{-1 } y, 
\end{equation} 
 
where V is $ N_{ sample } T \times N_q $ sized matrix with $ V_{ j k, l } = \bm{ \langle \Psi_j (k \delta t) \mid 0.5 (Z_l + 1) \mid \Psi_j (k \delta t) \rangle } $ and fixed teacher data $ y $. 
 
Then, $ N_{ sample } $, $ T $, $ N_q $, $ \Psi_j (k \delta t) $ and $ \delta t $ are the number of samples, number of time frames, number of qubits, state of quantum system at j-th sample at time k propagated by input data as the coefficient of operator $ 0.5 (1 + x_{ j k }) \mid 0 \rangle_0 \langle 0 \mid _0 + 0.5 (1-x_{ j k }) \mid 1 \rangle_0 \langle 1 \mid _0 $ and randomly made Ising or Floquet with Ising Hamiltonian, and time frame, respectively. 
 
Prediction result is $ \tilde { y } = V W $. 

The loss function of QRC is the squared difference between calculated and aimed waves. 
The bulk of QNN is the same as paper\cite{Wakaura_VQKAN_2024} and the input data is the coordinates of given function added into the coefficient of one body term on $ H_1 $ at first layer. 
The number of layers is 2 and the number of time frames is 10 for $ T $, respectively. 
The loss function of each point is calculated 10 times for QTCC. 
The coefficient of $ H_1 $ is the parameters for QNN and VQKAN.
The method for VQKAN is also same as the paper\cite{Wakaura_VQKAN_2024} and the parameter are made by the same manner as the paper\cite{Wakaura_VQKAN_2024}. 

The initial state is parabolic encoding $ \prod _{ j = 0 } ^{ N _q-1 } Ry^j (2 acos(\sqrt { _1 {\bf x}_j ^m })) \mid 0 \rangle ^{ \otimes N _q } $, respectively for each input $ m $.
For VQKAN, $ \phi_{ j k }^n (_n {\bf x}) = \sum_{ i \in \{ 0, dim (_n {\bf x}) \} }^{ in put ~ for ~ layer ~ n ~ N_d^n } acos (E_f (_n x_i) +\sum_{ s }^{ num. ~ of ~ grids ~ N_g }\sum_{ l }^{ num. ~ of ~ splines ~ N_s } c_s^{ n j k } B_l (_n x_i))$ and each $ \phi_{ j k }^n (_n {\bf x}) $ is the gate of the angle. 
 
Then, $ c_s^{ n j k } $ and $ B_l (_n x_i) $ are the parameter and spline functions at layer n, respectively, the same as classical KAN. $ _n {\bf x} $ is the input vector at layer n, $ j $ and $ k $ are the index of qubits, respectively. $ E_f (_n x_i) = _n x_i / (exp (-_n x_i) + 1) $ is the Fermi-Dirac expectation energy-like value of the distribution. The component of $ _n {\bf x } $ is the expectation value of the given observable for the calculated states of qubits. 
All calculations are performed in Jupyter notebook with Anaconda 3.9.12 and Intel Core i7-9750H.

 \section{Result of numerical simulations.}\label{3} 
\begin{figure} 
\includegraphics[scale=0.3]{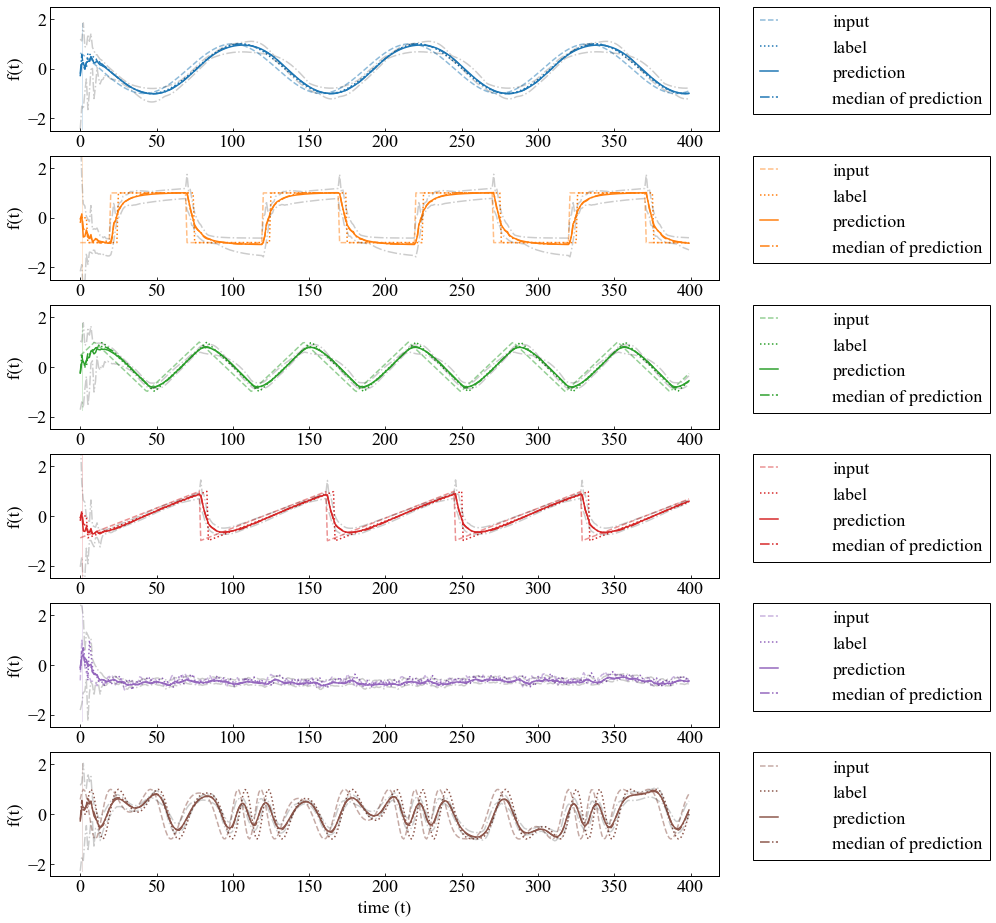} 
 \caption{ 
Time step v.s. the average, minimum and maximum of the result of generating an echo wave of sin, triangle, block, saw and two random waves, respectively for 10 attempts on QRC by noiseless time propagation. } 
\label{ q r c s } 
 \end{figure}

\begin{figure} 
\includegraphics[scale=0.3]{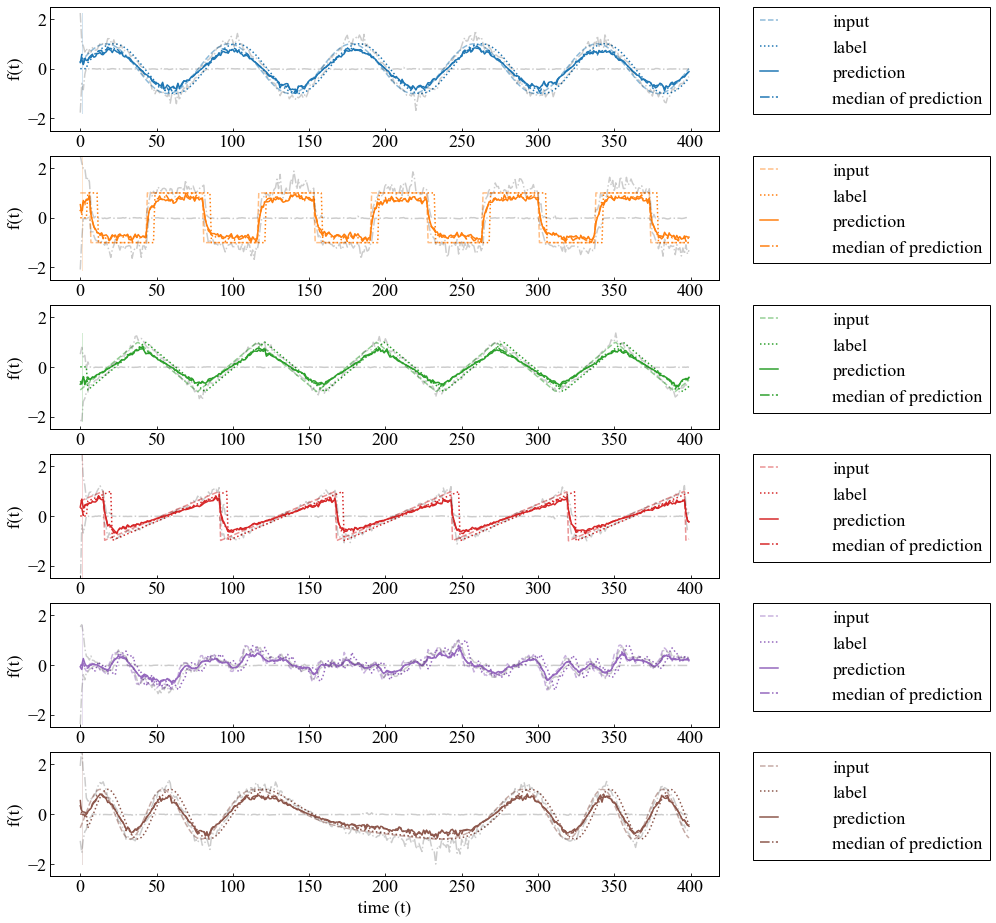} 
 \caption{ 
Time step v.s. the average, minimum and maximum of the result of generating an echo wave of sin, triangle, block, saw and two random waves, respectively for 10 attempts on QRC by QTCC. } 
\label{ q r c q } 
 \end{figure}

In this section, we show the result of demonstration of echo using QRC by QTCC, and fitting using QNN and VQKAN by QTCC, respectively. 
 The QTCC is not good at predicting the data using QRC as shown in Figs.\ref{ q r c s } and \ref{ q r c q }. 
The noise deviates the result of prediction in case of QTCC compared to that of noiseless time propagation. 
The average, minimum, and maximum of the result on QTCC are all larger than those of the result on noiseless time propagation, respectively as shown in Table.\ref{ q r c l }. 
The process of QRC should be adapted noises to take advantages of QTCC. 
 
\begin{table}[h] 
\caption{ 
The average, maximum and minimum of the calculated loss function of the result of generating an echo wave of sin, triangle, block, saw and two random waves, respectively, for 10 attempts by noiseless time propagation and QTCC, respectively.
 }\label{ q r c l } 
 \begin{center} 
 \begin{tabular}{c|c|c} \hline \hline 
 & noiseless& QTCC\\\hline 
 Average& 14.4402& 35.4215\\\hline 
Maximum& 26.4552& 92.84\\\hline 
Minimum & 9.3219& 22.5375\\\hline 
 \end{tabular}
 \end{center} \end{table} 
On the other hand, fitting using QNN and VQKAN by QTCC were more accurate than those of noiseless time propagation. 
 
 We describe the result of the fitting on equation (\ref{ last }) on 10 randomly sampled points, and the prediction of the values of given function on 50 sampled test points, which the optimizer is Covariance Matrix Adaptation Evolutionary Method (CMaEs) \cite{2016arXiv160400772H}, respectively. 
 
From this section, 
 
 \begin{equation} 
f^{ aim } ({\bf x}) =\exp\left({\sin}(x_0^2+x_1^2)+{\sin}(x_2^2+x_3 ^2)\right). \label{ last }
\end{equation}

\begin{figure} 
\includegraphics[scale=0.3]{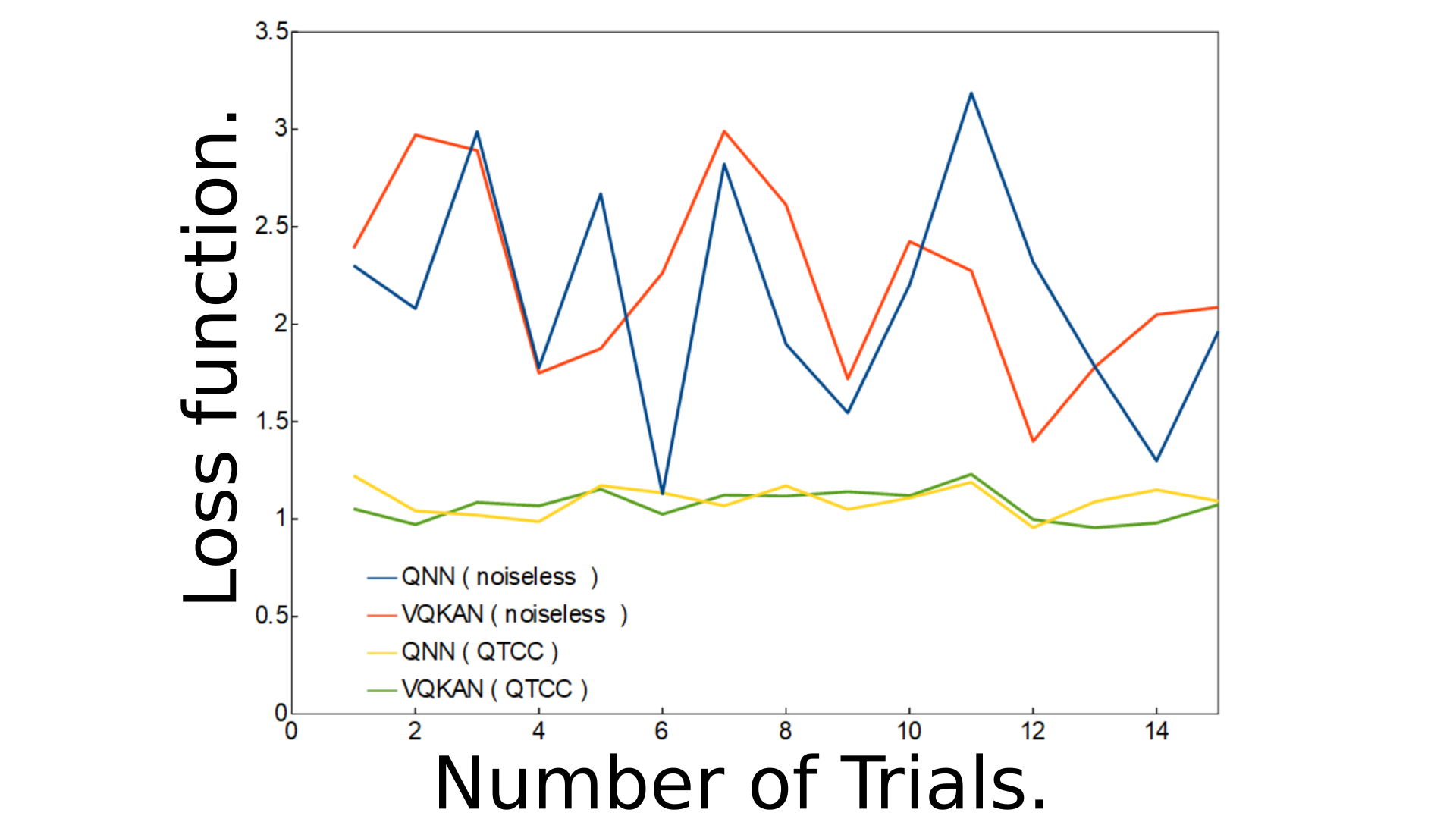} 
 \caption{ 
The number of trials v.s. the average of the value of Loss functions of the cases that (a) using QNN by noiseless time propagation, (b) using VQKAN by noiseless time propagation, (c) using QNN by QTCC, (d) using VQKAN by QTCC.} 
\label{ q t c c } 

 \end{figure}

 Here, $ x _ i = 2 _ 1 {\bf x} ^m _i-1 $ for $ i = 0, 1, 2, 3 $. $ _n {\bf x} ^m _i = 0.5 (\langle \tilde{ \Psi } (_1 {\bf x} ^m) | Z_i | \tilde{ \Psi } (_1 {\bf x} ^m) \rangle + 1)$ for the state calculated by n-th layer $ | \tilde{ \Psi } (_1 {\bf x} ^m) \rangle $, with $N_d^n = 4$ and $\dim(_n {\bf x} ^m) = 4$ for all layers and calculations, and the Hamiltonian is $Z_0 Z_1 + Z_2 Z_3$.
The realm of $ _n {\bf x} ^m$ is $ \{ 0, 0.25 \} $. 
We show the value of Loss functions of the cases that (a) using QNN by noiseless time propagation, (b) using VQKAN by noiseless time propagation, (c) using QNN by QTCC, (d) using VQKAN by QTCC in Fig. \ref{ q t c c }. 
The value of Loss functions of case (c) and (d) are smaller han those of case (a) and (b) at all trials. 
 The values of case (a) and (b) lower gradually, hence, the values may be smaller than those of case (c) and (d) after the trials. 
 
\begin{figure} 
\includegraphics[scale=0.3]{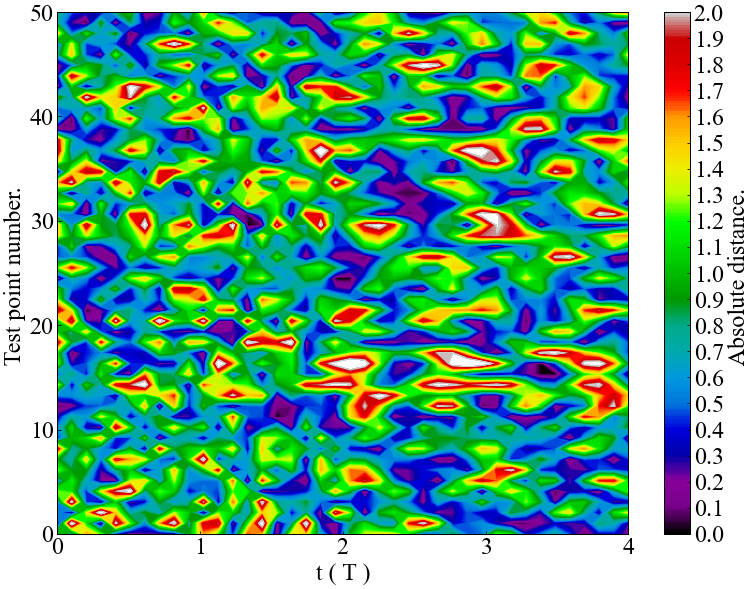} 
 \caption{ 
 The histogram of the average and of the sum of absolute distances between calculated and exact values (absolute distances) of test points for 10 attempts against time and the number of test point on the optimization using QNN by noiseless time propagation. } 
\label{ q t c c a t }  
 \end{figure}

\begin{figure}   
\includegraphics[scale=0.3]{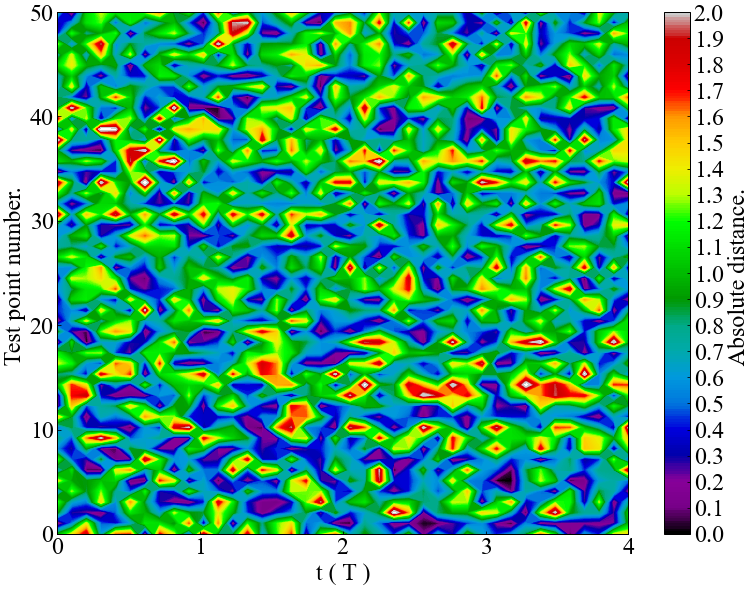} 
 \caption{  
 The histogram of the average and of the sum of absolute distances between calculated and exact values (absolute distances) of test points for 10 attempts against time and the number of test point on the optimization using VQKAN by noiseless time propagation. } 
 \label{ q t c c b t } 
 \end{figure}

In addition, we show the time propagation of the value of given function on 50 test points of case (a) in Fig. \ref{ q t c c a t } and (b) in Fig. \ref{ q t c c b t }, respectively. 
 
\begin{figure} 
\includegraphics[scale=0.3]{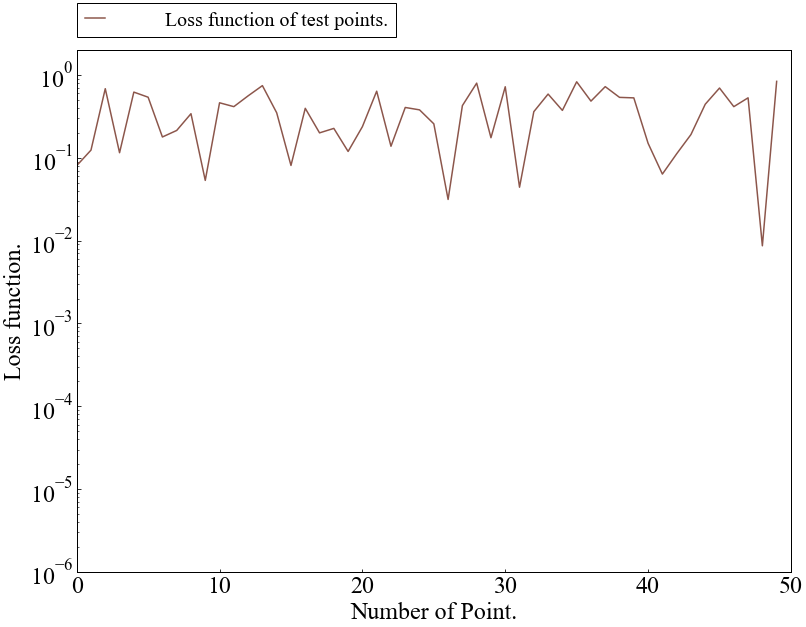} 
 \caption{ 
The number of test points v.s. the average and the median of the sum of absolute distances between calculated and exact values (absolute distances) of test points on the optimization using QNN by noiseless time propagation. } 
\label{ q t c c a p } 
 \end{figure}

\begin{figure} 
\includegraphics[scale=0.3]{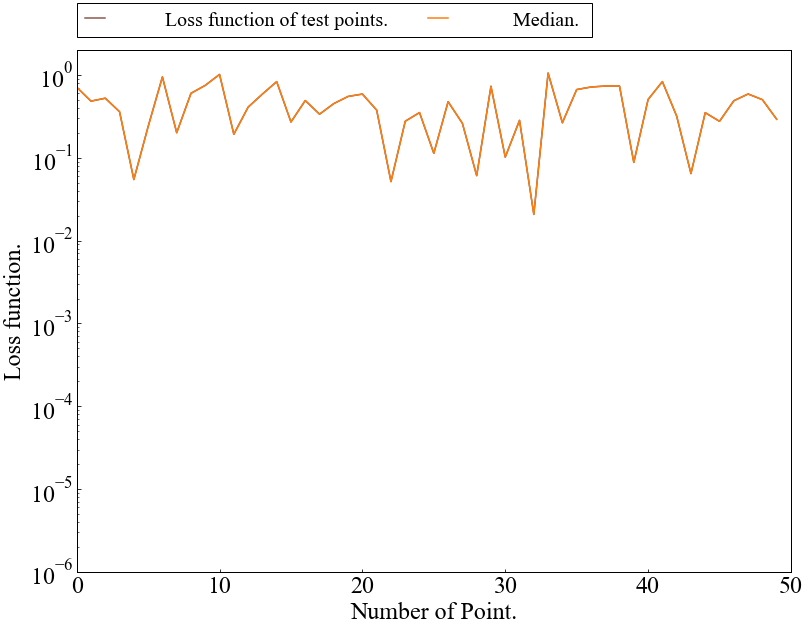} 
 \caption{ 
 The number of test points v.s. the average and the median of the sum of absolute distances between calculated and exact values (absolute distances) of test points on the optimization using VQKAN by noiseless time propagation. } 
\label{ q t c c b p } \end{figure}

 The value of absolute distances for time on each test point are unstably fluctuating for both cases. 
 Second, we show the result of prediction of the value of given function on 50 test points as the absolute distances between calculated and aimed value on 50 test points of case (a) in Fig. \ref{ q t c c a p } and (b) in Fig. \ref{ q t c c b p }, respectively. 
 
\begin{figure}  
\includegraphics[scale=0.3]{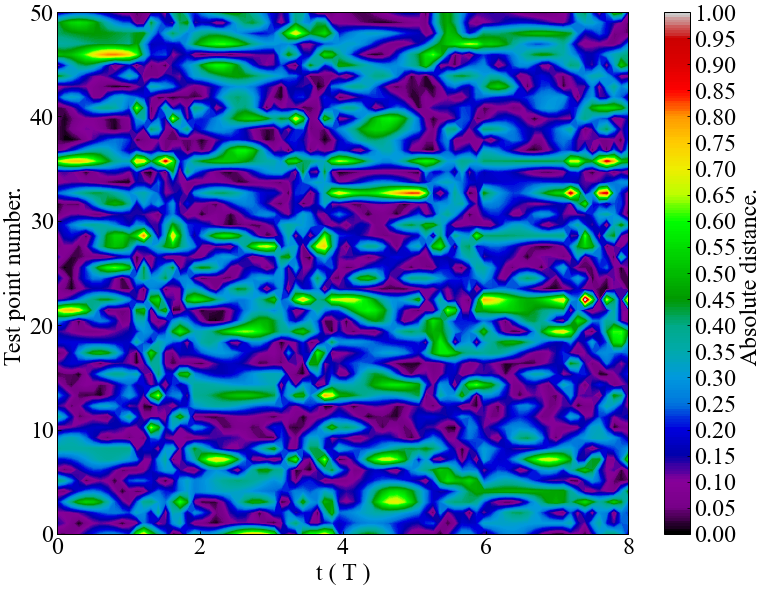} 
 \caption{  
The histogram of the average and of the sum of absolute distances between calculated and exact values (absolute distances) of test points for 10 attempts against time and the number of test point on the optimization using QNN by QTCC.} 
\label{ q t c c c t } 
 \end{figure}

\begin{figure} 
\includegraphics[scale=0.3]{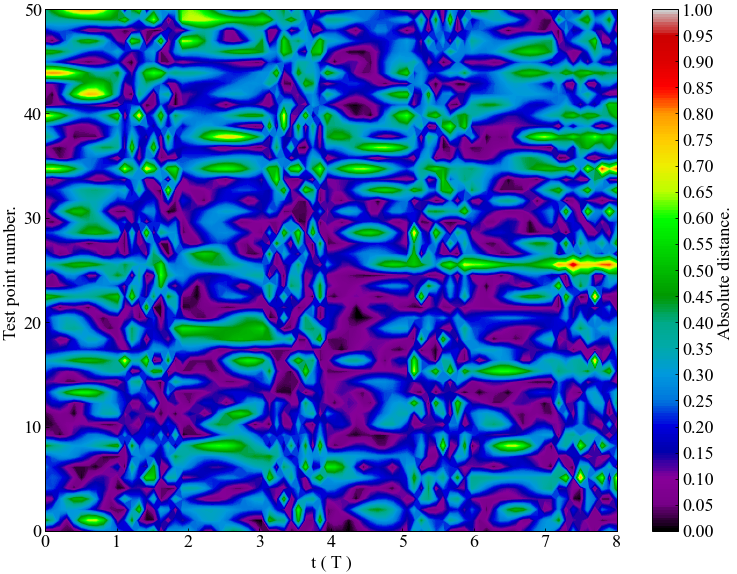} 
\caption{ 
The histogram of the average and of the sum of absolute distances between calculated and exact values (absolute distances) of test points for 10 attempts against time and the number of test point on the optimization using VQKAN by QTCC.} 
\label{ q t c c d t } 
 \end{figure}

\begin{figure} 
\includegraphics[scale=0.3]{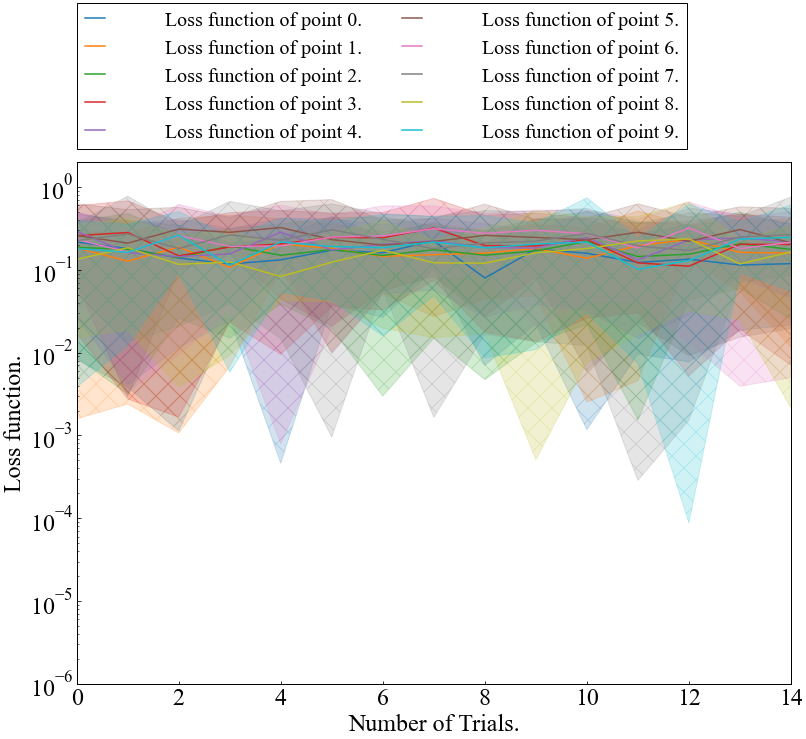} 
 \caption{ 
The number of trials v.s. the average of the sum of absolute distances between calculated and exact values (absolute distances) of sampled points for 10 attempts on the optimization using QNN by QTCC.} 
\label{ q t c c c i } 
 \end{figure}

\begin{figure} 
\includegraphics[scale=0.3]{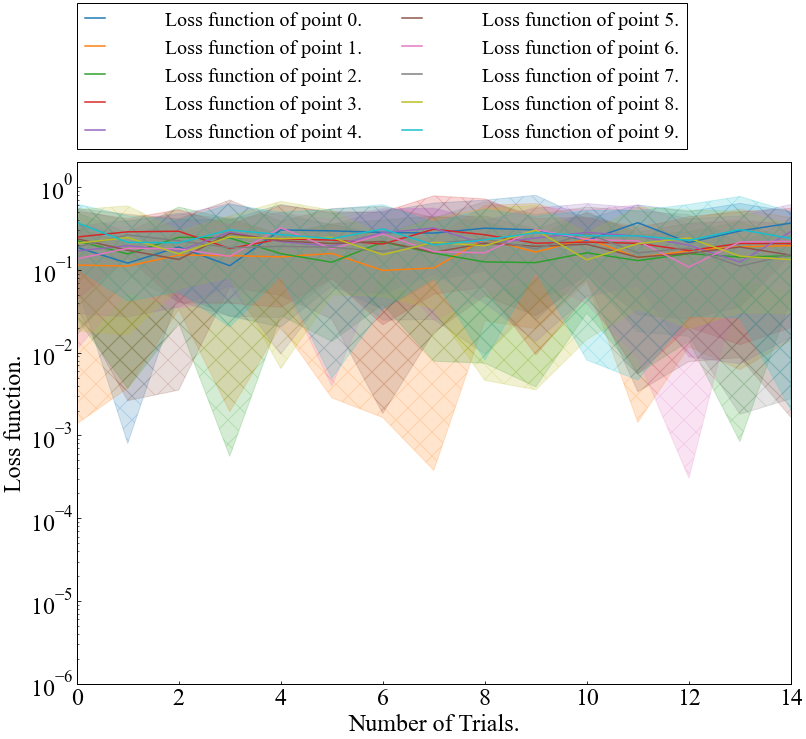} \caption{ 
The number of trials v.s. the average of the sum of absolute distances between calculated and exact values (absolute distances) of sampled points for 10 attempts on the optimization using VQKAN by QTCC.} 
\label{ q t c c d i } \end{figure}

 QNN is more accurate than VQKAN also according to Table.\ref{ q t c c l } with respect to the sum of absolute distances. 
It is supposed to be because the number of grids to interpolate the threshold function on gates as synapses is not enough large for only 15 trials. 
On the other hand, we show the time propagation of the value of given function on 50 test points of case (c) in Fig. \ref{ q t c c c t } and (d) in Fig. \ref{ q t c c d t }, respectively. 
The value of absolute distances for time on each test point are less fluctuated than those of case (a) and (b). 
According to Figs. \ref{ q t c c c i } and \ref{ q t c c d i }, the values of loss functions on each point are deviated for iterations of calculation due to the noise. 
 
\begin{figure} 
\includegraphics[scale=0.3]{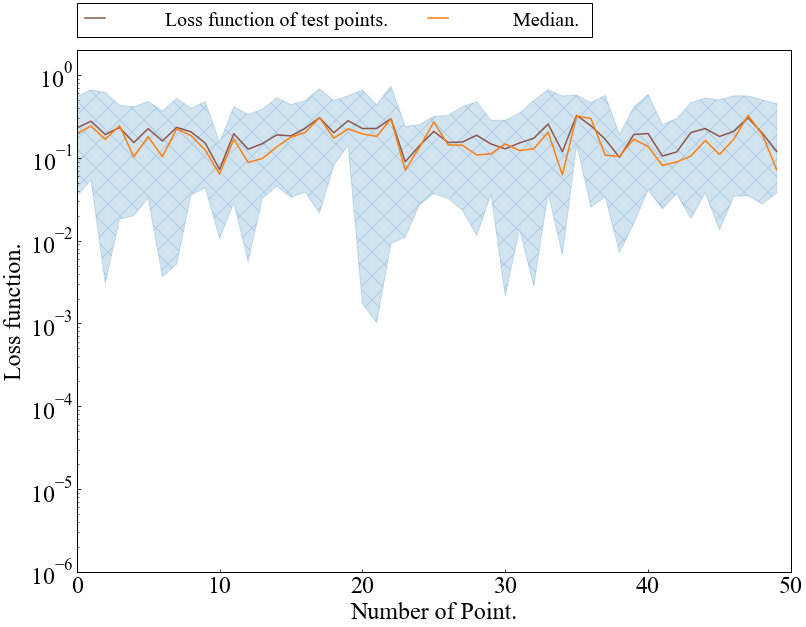} 
 \caption{ 
The number of test points v.s. the average and the median of the sum of absolute distances between calculated and exact values (absolute distances) of test points on the optimization using QNN by QTCC.} 
\label{ q t c c c p } 
 \end{figure}

\begin{figure} 
\includegraphics[scale=0.3]{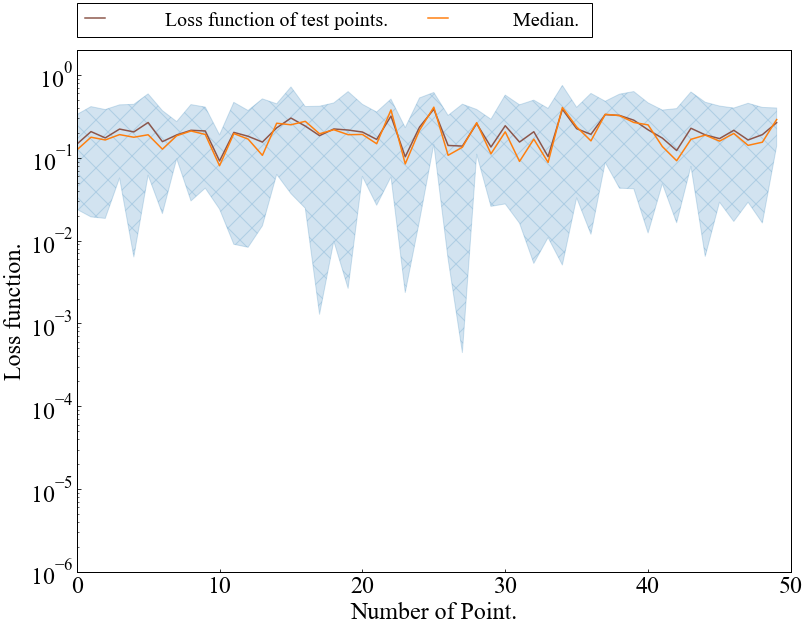} 
 \caption{ 
 The number of test points v.s. the average and the median of the sum of absolute distances between calculated and exact values (absolute distances) of test points on the optimization using VQKAN by QTCC.} 
 \label{ q t c c d p } 
 \end{figure}

Second, we show the result of prediction of the value of given function on 50 test points of case (c) in Fig. \ref{ q t c c c p } and (d) in Fig. \ref{ q t c c d p }, respectively.
 
\begin{table}[h] 
\caption{ 
The sum of the absolute distances on the result of prediction of function values (a) using QNN by noiseless time propagation, (b) using VQKAN by noiseless time propagation, (c) using QNN by QTCC, (d) using VQKAN by QTCC. 
 }\label{ q t c c l } 
 \begin{center} 
 \begin{tabular}{c|c} \hline \hline 
 & absolute distances \\\hline
QNN (noiseless) & 18.6153\\\hline 
VQKAN (noiseless)& 22.1747\\\hline 
QNN (QTCC) & 9.5243\\\hline 
 VQKAN (QTCC) & 10.4943\\\hline 
 \end{tabular} 
 \end{center} 
\end{table}

 QNN is more accurate than VQKAN also according to Table.\ref{ q t c c l } with respect to the sum of absolute distances same as the case (a) and (b), and the values themselves are much smaller than those of them. 
The QTCC improves the accuracy of prediction because the noise on Hamiltonian of TC makes the values of calculated loss functions of each point the distribution, hence, the sum of averages may be smaller than the sum of the values of calculated loss functions of noiseless time propagation. 
 
\section{Concluding remarks} \label{4} 
 
In this paper, we revealed that QTCC can be used in quantum machine learning, and improves the accuracy of prediction. 
It means that rotational noise may contribute to the accuracy of quantum machine learning. 
However, it is the only result of solving one problem, thus, solving other problems and surveying the effect of QTCC are the next problems. 
Besides, natural QTCC may exist in unexpected systems such as neurons 
\cite{toida_magnetometry_2023}. 
Surveying them is also next problem.

 \newpage\bibliographystyle{apsrev4-2}

\bibliography{main}

\end{document}